# VisiSploit: An Optical Covert-Channel to Leak Data through an Air-Gap

Mordechai Guri, Ofer Hasson, Gabi Kedma, Yuval Elovici

Ben-Gurion University of the Negev

{gurim, hassonofer, gabik, elovici}@post.bgu.ac.il

**Abstract**

In recent years, various out-of-band covert channels have been proposed that demonstrate the feasibility of leaking data out of computers without the need for network connectivity. The methods proposed have been based on different type of electromagnetic, acoustic, and thermal emissions. However, *optical* channels have largely been considered less covert: because they are visible to the human eye and hence can be detected, they have received less attention from researchers.

In this paper, we introduce VisiSploit, a new type of optical covert channel which, unlike other optical methods, is also stealthy. Our method exploits the limitations of human visual perception in order to unobtrusively leak data through a standard computer LCD display. Our experiments show that very low contrast or fast flickering images which are invisible to human subjects, can be recovered from photos taken by a camera. Consequentially, we show that malicious code on a compromised computer can obtain sensitive data (e.g., images, encryption keys, passwords), and project it onto a computer LCD screen, invisible and unbeknownst to users, allowing an attacker to reconstruct the data using a photo taken by a nearby (possibly hidden) camera. In order to demonstrate the feasibility of this type of attack and evaluate the channel's stealth, we conducted a battery of tests with 40 human subjects. We also examined the channel's boundaries under various parameters, with different types of encoded objects, at several distances, and using several kinds of cameras. Our results show that binary data can be leaked via our covert channel. Further research and discussion may widen the scope of this field beyond its current boundaries, yielding novel attack paradigms that exploit the subtle mechanisms of human visual perception.

## 1. Introduction

Despite a widening array of defensive measures - including intrusion detection and prevention systems (IDS/IPS), firewalls, antivirus programs and the like - persistent attackers continuously find innovative paths to infiltrate malicious code into target systems. Even so-called 'air-gapped' systems can be breached or bypassed by skillful combinations of devious malware and attack vectors [1] [2] [3] [4] [5].

While breaching such systems has been shown feasible in recent years, exfiltration of data from systems without networking or physical access is still considered a challenging task. Electromagnetic radiation of communication networks, keyboards, and computers is one of the first type of covert channel researched [6] [7] [8] [9]. Other types of acoustic [10] [11] and thermal [12] out-of-band exfiltration channels have also been suggested. A few optical methods have been presented as well [13] [14]. However, these methods, which rely on their optical visibility, presume the absence of people in the environment (e.g., after the work day in the evening hours).

In this paper, we propose an optical covert channel, codenamed VisiSploit, which relies on the limitations of human vision. Technically speaking, visible light represents a limited range of electromagnetic radiation (with wavelengths between 390 to 700 nm), which is sensed and perceived by the human visual system. Intentional leakage of sensitive data through the emanation of visible light via a standard LCD screen is futile, since by definition it may be detected by humans who see the display. Our covert channel exploits the limits of human visual perception in order to conceal images (encoded with sensitive data), invisible to the naked eye, on the computer screen. Such images can be acquired by a nearby digital camera and reconstructed through digital image processing.

The contribution of this work is of threefold:

1. To the best of our knowledge this is the first optical covert channel which uses a computer display screen to exfiltrate sensitive data over an air-gap. Compared to existing optical methods which demand user absence, our method is covert, and hence can be used while the user is present or working on his/her computer.
2. We conducted an experiment with 40 human subjects in order to evaluate the stealth of our method and determine the optimal threshold for image concealment.
3. We performed an extensive set of experiments with different types of graphical objects, cameras, and distances to determine the effective distance at which the leaked image can be reconstructed and the quality of the reconstructed image.

The rest of the paper is organized as following. In Section 2 we present related work. Section 3 describes the threat model. Section 4 discusses architecture and implementation. Section 5 presents tests and evaluation results. Section 6 provides brief scientific background. Section 7 discusses countermeasures, and we conclude and mention future work in Section 8.

## 2. Related work

Exfiltrating information over an air-gap via covert channels has been the subject of research over the years. Covert channels investigated are either electromagnetic, acoustic, thermal, or optical. Kuhn and Anderson [6] discuss 'soft tempest' involving hidden data transmission using electromagnetic emanation. Funthenna [15] and AirHopper [16] are attack patterns aimed at exfiltrating data from air-gapped networks using generated radio frequency emanations. Other electromagnetic and magnetic covert channels are discussed in [17] [9] [18] [19]. Hanspach and Goetz [20] present a method for near-ultrasonic covert networking using speakers and microphones. Fansmitter is another method of acoustic data exfiltration from computers with no speakers [21]. BitWhisper [22] demonstrates a covert communication channel between adjacent air-gapped computers by using their heat emissions and built-in thermal sensors.

### 2.1 Optical methods and shoulder surfing

In the optical domain, Loughry and Umphress [13] discuss the risks of information leakage through optical emanations from computer LEDs. They implemented malware that manipulates the keyboard LEDs to transfer data to a remote camera. A unique infiltration method proposed by Shamir et al demonstrated how to establish a covert channel with a malware over the air-gap using a blinking laser and standard all-in-one-printer [14]. However, these optical methods assume that the user is absent from the room, in order to avoid detection which may take place when the user is present. In 2013, Brasspup [23] demonstrates how to conceal secret images in a modified LCD screen. His method required officiation of the LCD screen (removing the polarization filter) which makes it less feasible in an attack model. The concept of shoulder surfing [24] is discussed by Lashkari et al [25], and Kumar et al [26]. The classic threat of shoulder surfing refers to a malicious insider or visitor (or an exploited surveillance camera [27]) obtaining confidential data such as a password or PIN, as the legitimate user enters the data. With our method, the presence or absence of the legitimate user is not required, since the malware may covertly project the sensitive information as needed, at any time. Table 1 summarizes the different types of existing air-gap covert channels, and presents their effective distances.

**Table 1. Different type of air gap covert channels and distances**

| Method | Examples | Effective Distance |
|---|---|---|
| **Electromagnetic** | [6] [7] [9] [15] [17] [9] [16] | ~5-10M |
| **Acoustic** | [11] [10] [21] | ~15M |
| **Thermal** | [12] | 40cm |
| **Optical** | [13] [14] [23] Shoulder Surf [25] [26] | Line of sight |
| Optical | VisiSploit | ~1-8M |

### 2.2 Optical Covert-Channel vs. Steganography

Steganography is the art of concealing data within other data (e.g., documents, media streams, or communication protocols). Images are the most popular objects used for steganography, and image steganography techniques may include changing pixel intensities, exploiting characteristics of image compression algorithms, using color redundancy, and more. A comprehensive overview of image steganography is given in [28].

Image steganography techniques are not aimed at leaking data through an air-gap. In an air-gapped case, the image is captured *optically* as a snapshot by a remote camera, hence many of its visual properties are not preserved (e.g., image resolution and quality, color intensities, etc.). Consequently, since the original image differs from the new one, usually encoded message is not recoverable. To the best of our knowledge no work has been conducted in the steganography field investigating the application of steganography in an air-gap scenario. Our work focuses on an *optical* covert channel which can be used to reconstruct messages concealed within images taken by an external camera, despite the optical limitations. In particular, we conduct comprehensive tests with different types of cameras, from a number of distances, and at various thresholds. We also measured the distance at which the data is recoverable and the rate. The differences between traditional image steganography and an optical covert channel are presented in Table 2.

**Table 2. Differences between image steganography and an optical covert channel**

|  | Concealed from | Reconstructed from |
|---|---|---|
| Image steganography | Human eye | Original image (transferred digitally) |
| Optical covert-channel | Human eye | Image taken from remote camera (loss in resolution and image quality) |

## 3. Threat model

The threat model necessitates contaminating the target computer with a crafty malicious program. Infecting a computer within a secured network can be accomplished, as demonstrated in recent years [29] [30] [31] [32] [33] [34] [1]. The malicious code gathers sensitive information from the user's computer (e.g., password, encryption keys), encodes it as a visual object (e.g., QR code), and projects it on the computer's display, using visual effects that conceal it from human perception. The threat model also requires a digital camera which can take single shots or video recordings of the compromised computer's display. Attackers can then reconstruct the sensitive information by applying image processing techniques on the images or video obtained.

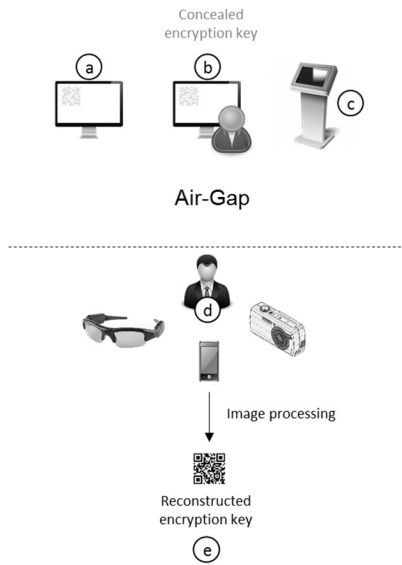

**Figure 1. Illustration of the optical covert channel.**

In particular, we identify two main scenarios in which the threat model is most relevant. (1) The 'Evil maid' [35] or 'malicious insider' [36] type of scenario, in which a person with a camera can be within the compromised computer's line of sight but not necessarily have direct network access. (2) A scenario involving a compromised network which has one or more display screens exposed to the public domain, within the line of sight of a person with camera (e.g., dashboards, kiosk screens, ATM machines, etc.). The covert channel is depicted in Figure 1. Sensitive data (encryption key) is encoded as a quick response (QR) code and covertly concealed on the computer screen. It can be projected on the screen either when the user is absent (a) or when the user is working on the computer (b). Alternatively, it can be concealed on a publicly accessed screen (e.g., dashboard), if available (c). A person in close proximity to the computer (d), equipped with a camera (e.g., hidden camera) takes a photo or video of the computer screen. The concealed data is then reconstructed from the photographic or video images using image processing techniques (e) (Figure 2).

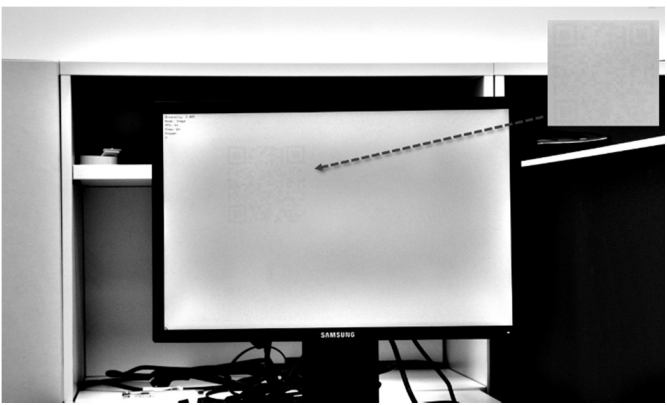

**Figure 2. Sample photo of screen (with concealed QR), *after* basic image processing.**

Notably, threat models in which the attacker must be in close proximity of the emanated device are common in a variety of covert channels [13] [25] [16] [9].

## 4. Design and implementation

In this section we describe the basic concealment methods and describe the implementation of our optical covert channel.

### 4.1 Objects

During our experiments, we tested three kinds of visual objects (Figure 3): an image, a QR code, and plain text. The three types of objects reflect different types of data that an attacker might want to exfiltrate using the covert channel. Images represent pictures, drawings, computer aided design (CAD) graphics, and so on. While the original image might not be recovered perfectly, the main outline is constructible so it remains potentially valuable. We specifically use an image of an office building plan (scaled from an original 1400x906 bitmap). The image outlines 14 offices, three meeting rooms, a lobby, a kitchen area, and a bathroom. The image also includes detailed information such as the locations of 63 workstations, tables, and chairs. QR code, another type of visual object, is used during the experiments. The QR code [37] represents two-dimensional barcode which consists of white and black dots. The amount of data depends on: (1) the encoding scheme, (2) the version, and (3) the error correction level. The encoding scheme determines whether QR stores numeric/alphanumeric characters or binary data. The version is the dimension of the symbol and is the main factor in QR storage capacity. Based on the Reed-Solomon algorithm, QR codes have five levels of error correction: level L (in which 7% of the code words can be recovered), level M (15%), level Q (25%), and level H (30%) [37]; the higher the error correction level, the less storage capacity. QR is particularly interesting for VisiSploit, since it allows exfiltration of textual and binary information which are also error-sensitive. For our experiments, we use version 3 QR code (29x29) with error correction level M, which can encode 352 bits of data (101 digit number). We also tested our method with simple plain text images of three different sizes. This type of image is suitable for leaking short textual data such as passwords or identifiers.

**Figure 3. The image, QR code, and plain text objects used for the evaluation of our covert channel.**

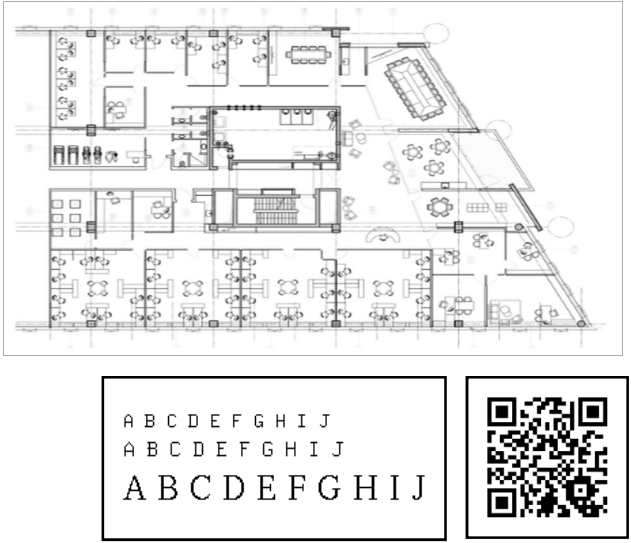

## 4.2 Image concealment

It is known that human vision faces limitations resolving fast flickering images and certain color differences and wavelengths [38] [39] [40] [41]. For interested readers, we present the scientific background on this subject in Section 6. We use two techniques to conceal images in the background: (1) embedding the image on a bright or dark background surface ("embedding"), and (2) blinking the image at high frequency ("blinking").

### *4.2.1 Embedding image on bright/dark surface*

Modern computer monitors are capable of displaying each pixel in any one of at least 16,777,216 color variations. Usually, the color of each pixel is defined as triples of red, green, and blue (RGB) channels. Each channel is encoded by eight bits, which represent the intensity of the component in the triple. This scheme is referred to as true color or 24-bit color depth. Figure 4 depicts three samples of embedding images on bright and dark surfaces with close RGB. The nine circles have an increasing intensity from left to right. Obviously, where the brightness value of the object is equal to the brightness value of the background (left circle), the object becomes invisible. Where the brightness value of the object is close to the brightness value of the background (right eight circles) it may be visually difficult to notice and detect some the objects by the naked eye.

**Figure 4. Samples of bright on bright and dark on dark surfaces with increasing contrast.**

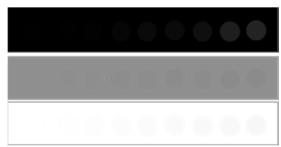

### *4.2.2 Blinking image*

The refresh rate is the number of times in one second that the screen draws the displayed image. Older LCD monitors have a refresh rate of 60Hz, while newer monitors have a rate of 120Hz or more. The number of frames the computer can change per second is called frames per seconds (fps), and this is a dependent on both the video card and the computer screen. 60 fps is the default setting of many desktop computers. By only displaying a bitmap in a few frames every second, the effect is a fast blinking image which is difficult for the human eye to perceive. When the image exposure time falls below a given threshold, the human visual system will fail to detect the object. We also combined the two techniques by using the intensity (previously discussed in subsection 4.2.1) and flashing together.

## 4.3 Concealment process

Concealing an image on a background involves three main steps: (1) locating a bright or dark surface to embed the image in, (2) histogram compression: reducing the image to a binary (two color) image at a specified level of low intensity, and (3) embedding the image in the surface.

For testing, we implemented a VisiSploit prototype for Windows OS. Our implementation consists of the image concealment module (Figure 5) which is a dynamic-link library (DLL) that mediates between the application and its Graphics Device Interface (GDI) DLL. The GDI itself interacts with device drivers and handles graphical operations such as drawing bitmaps and text, filling shapes, and more. The image concealment module is capable of intercepting and modifying different Windows OS GDI functions. In our prototype, we locate bright or dark *regions* and draw the image objects into it. More on GDI regions, drawings and clipping can be found in [42]. Note that the concealment module must be injected into each process separately. System wide implementation would require the construction of a kernel level display driver.

**Figure 5. VisiSploit's image concealment module**

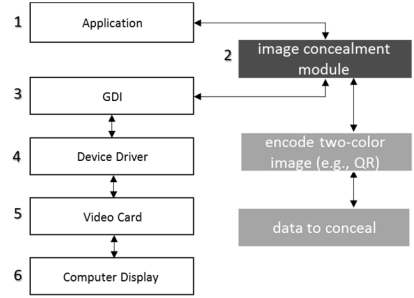

**Reduction to two color image.** Reduction of the image to a two color image involves an intermediate step of transferring it to a grayscale image [43]. To create a two color image (from a grayscale image), at a specified intensity level, we use a cluster based threshold using Otsu's algorithm [44]. Note that the contrast levels are represented in this work in

| | | QR | | | Text | | | Image | | |
|---|---|---|---|---|---|---|---|---|---|---|
| | | Minimal | Average of 40 subjects | Stdev | Minimal | Average of 40 subjects | Stdev | Minimal | Average of 40 subjects | Stdev |
| Bright | Static | 2.11% | 2.45% | 0.39% | 3.50% | 4.46% | 0.49% | 1.83% | 3.18% | 0.93% |
| | Blink 60 | 3.68% | 5.79% | 1.22% | 12.00% | 16.03% | 3.27% | 4.26% | 6.36% | 1.09% |
| | Blink 30 | 4.08% | 5.61% | 0.79% | 13.50% | 15.88% | 1.99% | 4.74% | 5.98% | 0.72% |
| Dark | Static | 0.20% | 0.36% | 0.19% | 2.50% | 4.53% | 1.57% | 0.15% | 0.94% | 0.70% |
| | Blink 60 | 0.20% | 0.86% | 0.50% | 6.00% | 13.80% | 3.91% | 0.68% | 1.52% | 0.63% |
| | Blink 30 | 0.39% | 0.92% | 0.51% | 6.50% | 14.23% | 3.63% | 0.49% | 1.58% | 0.64% |

Table 3. The threshold results for 40 subjects tested

the form of the ratio between background and image pixels calculated using the following formula:

$$contrast\% = 100(\frac{|P_b - P_i|}{255})$$

Where $P_b$ and $P_i$ are the grayscale values of the background surface and the image.

### 4.4 Reconstruction process

After taking a photo or video of a monitor, we use image processing in order to reconstruct the concealed object. we provide a brief description of the reconstruction process here, while a more detailed and mathematical explanation is provided in Appendix A. This process consists of three steps: (1) image desaturation, (2) dynamic range extension, and (3) image sharpening. Desaturating the image to transform it from sRGB color space (standard RGB) into grayscale color space. This step removes background noise and colored pixels at unnecessary wavelengths. Extension of the dynamic range by around 40-50% is achieved by using histogram equalization [45]. This greatly increases the contrast and significantly helps in the identification of the object and its reconstruction. Final sharpening of the contrasted image is achieved using the unsharp masking method [46]. Automatic and improved image reconstruction might involve computational analysis and advanced computer vision methods, however this is left for future work.

### 5. Tests and evaluation

The initial phase of evaluation aims to reveal the stealth threshold, i.e., the minimal amount of contrast intensity between an object and its background needed, whereby human subjects can still notice the presence of an object while working their computers. We evaluated the threshold of visibility by conducting a series of tests with human subjects. At the beginning of the tests, subjects were told about the existence of a hidden object (image, QR, or text) in white and black surfaces. We control the intensity of the object and its blinking frequency, while the human subject staring right at the computer's display. When the subject detects the image on the screen, the tester makes note of the threshold's value, and the subject proceeds to the next image. Note that the subject is seated at a comfortable distance and angle from the display, when he/she *knows* in advance about the existence of an object on the display. In a real covert channel scenario, the computer user is not aware of the presence of an object or when it is displayed, and he/she does not know the object's position on the display. Consequently, we assume that the threshold values obtained by our in-lab tests should be more than adequate for a real-life scenario, at least in terms of the user's awareness.

### 5.1 Concealment results

Table 3 presents the testing results of the 40 subjects tested, for QR, text, and image tests, each in bright and dark configurations. The measurements are given in thresholds, which represent the contrast level between the object and its white or black background (a lower threshold means lower contrast). As noted, we use percentage values instead of numerical color values, which is justified, since we are interested in the perceived contrast between the intensity of the object and its background.

Table 3 includes the minimal brightness/darkness threshold noticed by the subjects. It also includes the average of this threshold over the 40 subjects, and the standard deviation. The major configurations are 'Bright' (bright object on white background) and 'Dark (dark object on black background). These configurations are further separated into 'Static,' 'Blink 60 Hz,' and 'Blink 30 Hz.' The values, in percent, refer to the brightness of the object and the background. Each subject was requested to indicate the point at which he/she could see the object (the minimal delta or contrast).

As can be seen, the average and minimal thresholds of the 'Dark on Dark' sub-configurations are, in general, significantly lower than the respective values of the 'Bright on Bright' sub-configurations. Effectively, this means that embedding images on a white background is a better choice for the optical covert channel. We note that prior to the test, subjects were informed about the existence of a hidden object in the display. We assume that the thresholds are even higher in a real-life scenario, when user is not aware of an image concealed on the screen. Note that the average values of the text object pose an exception. Its thresholds are higher than those of the QR and image objects, meaning that the hidden text images were not well identified, compared to the concealed image and QR objects This phenomena is due to the relatively small emission surface (small number of pixels) compared to QR and image objects.

## 5.2 Reconstruction results

Equipped with the thresholds for object concealment, we conducted a set of experiments and tried to reconstruct the concealed images from photos taken by remote cameras. During the experiments we used different camera types, several distances, and various data encoding patterns and color schemes. Still photographs were used in the static image tests, while video recordings were used in the blinking image tests.

The contrast values between the object and the background intensities were derived from the results of our earlier experiments with human subjects. We used contrast values slightly below the minimal contrast values perceived by humans. Consequently, we have comfortable (larger) 'safety margins.'

The images were manually reconstructed by digital image processing tools, until the maximal feasible resolution was achieved. We used the open source GIMP tool [47] to perform the image desaturation, dynamic range extension, and image sharpening. With the video recordings used in the blinking image experiment variant, additional preprocessing was required: scanning the sequence of frames to find those frames that actually capture the object, among the empty frames. We focused primarily on three types of objects: images (the picture of an office building plan), data encoded in QR code, and plain text.

## 5.3 Experimental setup

The LCD displays used throughout the experiments were the off-the-shelf 22" Samsung SyncMaster 2243BW and S24D590PL 23.6" which were connected to a desktop computers. The display screen settings such as brightness, contrast, position, sharpness, and color balance remained at their default values. We conducted the experiments during the daylight hours in a real workspace, with fair exposure to light that simulates a regular office environment. The overall settings, including the level of ambient light, the computer's display type, etc. were the same settings used during the experiments with human subjects.

### 5.3.1 Camera types

During the tests, we used five types of cameras: entry-level and professional-level Digital single-lens reflex (DSLR) camera (with 35mm and 135mm lenses), an extreme camera, a high definition (HD) webcam, a smartphone camera, and a wearable camera. Relevant technical specs are provided in Table 4. Note that the types of cameras we selected reflect different attacks scenarios. The photo or video can be taken by a high quality camera (i.e., an insider or visitor), webcam connected to a computer infected by a malware, a smartphone with a malicious spy application, or by visitor with a wearable camera device (such as Google Glass and "Spy Watch").

**Table 4. Type of cameras tested**

| Type | Model | Resolution | Video |
|---|---|---|---|
| **Still camera (entry-level DSLR)** | Nikon D7100. lens: Nikon18-140mm F3.5-5.6 ED VR | 6000x4000 1920x1080 (video) 1280x720 (60 fps video) | 30-60 fps |
| **Still camera (professional-level DSLR)** | Canon 5D Mark III full-frame | 5760x3840 | - |
| **Still camera (extreme)** | GoPro Hero4 | 4K - WVGA | 24-120/240 fps |
| **Webcam (HD)** | Microsoft LifeCam Studio | 1920x1080 1280x720 (video) | 15-30 fps |
| **Smartphone camera** | Samsung Galaxy S5 (I9505) | 4128x3096 1920x1080 (video) | 30 fps |
| **Wearable device** | Google Glass Explorer Edition | 2528x1856 1280x 720 (video) | 30 fps |

## 5.4 Tests and results

During testing we took photographs from a direct angle and short 'shoulder surfing' range of 50-100cm to a distance of 8m. Our tests reveal that it is possible to reconstruct visual objects secretly displayed on a computer monitor. The best results were achieved with a static image on a bright background, using data encoded as QR code and a DSLR camera. This is particularly useful, since QR code also provides a convenient way to encode encryption keys using Base64. Table 5 (QR code), Table 6 (text) and Table 7 (image) summarize the results of static reconstruction at different distances using a DSLR camera.

**Table 5. Reconstruction rates for white on white and dark on dark QR 29x29 with a DLSR camera (35mm and 135mm lenses)**

|  | QR (bright) | QR (dark) |
|---|---|---|
| **50cm** (DSLR - 35mm) | 95% | 95% |
| **1m** | 75% | 75% |
| **2m** | 70% | 75% |
| **3m** (DSLR - 135mm) | 80% | 75% |
| **4m** | 75% | 70% |
| **5m** | 80% | 70% |
| **6m** | 70% | 40% |
| **7m** | 75% | 30% |
| **8m** | 70% | 10% |

**Table 6 Reconstruction rates for white on white and dark on dark text with a DLSR camera (35mm and 135mm lenses)**

|  | Text (bright) | Text (dark) |
| --- | --- | --- |
| **50cm (DSLR - 35mm)** | 100% | 100% |
| **1m** | 100% | 100% |
| **2m** | 50% | 80% |
| **3m (DSLR - 135mm)** | 20% | 10% |

**Table 7. Reconstruction rates for white on white and dark on dark image with a DLSR camera (35mm and 135mm lenses)**

|  | Image (bright) | Image (dark) |
| --- | --- | --- |
| **50cm (DSLR - 35mm)** | 100% | 80% |
| **1m** | 60% | 60% |
| **2m** | 30% | 30% |
| **3m (DSLR - 135mm)** | 10% | 70% |
| **4m** | 10% | 70% |
| **5m** | 10% | 50% |

As can be seen in the white on white case, QR code has the highest reconstruction level. We were able to reconstruct 70-100% of the encoded QR code at a distance range of 50cm to 8m. Text was reconstructable at 4m and images at a maximum range of 2m. Dark on dark images were significantly more difficult to reconstruct, and in this case 70% of the QR code was reconstructed, when the photographs and video was taken from a distance of 4-5m; text and images in this scenario were reconstructed at a distance of at most 3m. We identify QR as the best candidate for reconstruction as it consists of small rectangles of a known size, which are well-defined in the 2D space. Text and image objects are more difficult to reconstruct, since they consist of less well-defined regions and connectors.

Sample photos taken during testing are shown in Figure 6. To the naked eye, the original image reveals no information aside from the software parameters displayed in the corner (included for ease of record keeping). Following image processing, the hidden QR code is reconstructed. The parameters were also tested using human subjects looking directly at the screen (as opposed to the captured image) so as to simulate an unsuspecting user working on the computer.

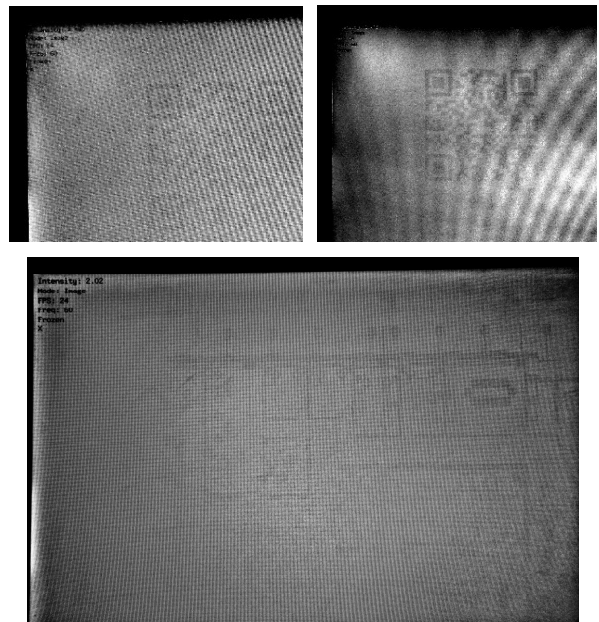

**Figure 6. Sample of photos taken during testing, following basic image processing.**

### 5.4.1 Professional DSLR

To improve the dark on dark results we went on to test enhanced equipment and settings, probing for optimal results. The camera used was a Canon 5D Mark III, with the following lenses: (a) EF 24-70mm f/2.8L II USM; and (b) 70-200mm F/4 IS USM. We used a tripod to stabilize the camera. We tested QR under the dark on dark settings using this camera, a scenario in which the entry level DSLR showed limited results.

With the 24-70mm lens, reconstruction was 100% at distances of 50cm-200cm. With the 70-200mm lens, reconstruction was 80-100% at distances of 3-8m. As can be seen, the results with the professional equipment are significantly better than those of the basic one.

### 5.4.2 Extreme (GoPro), Smartphone, Webcam, and Wearable Cameras

Results shows that the smartphone camera is limited to a short distance of at most 1 meter. Limited results for short distances were also achieved using the webcam. Table 8 summarizes the results of reconstructing white on white blinking images at 30Hz from a video taken by GoPro, DSLR camera (Nikon), smartphone, LifeCam, Google Glass. We have used rectangles of three sizes; 100x100 (small), 200x200 (medium), and 300x300 (large) pixels.

**Table 8. Reconstruction rates for white on white blinking rectangle with GoPro, DSLR, Webcam, and Wearable cameras**

| Bright | GoPro | DSLR - 35mm | Smart-phone | LifeCam | Google Glass |
| --- | --- | --- | --- | --- | --- |
| **50cm** | Small | Small | Small | Small | Small |
| **100cm** | Small | Small | Small | Small | Large |
| **150cm** | Small | Small | Large | Small | - |
| **200cm** | Small | Small | - | Medium | - |
| **250cm** | Small | Medium | - | Large | - |
| **300cm** | Small | Large | - | | - |

*5.4.3 Wearable camera (Google Glass)*

Google Glass and some types of wearable hidden cameras (e.g., "spy pen" [48]) have a low quality-level of a 5MP camera. We were able to recover about 30% of a large QR code from a distance of 50cm. This distance closely simulates an unsuspecting user wearing the device while standing in front of the hidden object rendered on monitor and captured by the camera. A small portion of the recovered QR code is shown in Figure 7. We were also able to identify the blinking block from a limited range of 50-100cm. More encouraging is the fact that object recovery rates are much better on bright backgrounds than dark backgrounds. This is beneficial to attackers, since objects on bright backgrounds are much easier to hide from the naked eye.

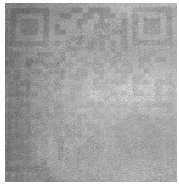

**Figure 7. QR code reconstructed from a photo taken with Google Glass from a distance of 50cm.**

*5.4.4 Dashboard scenario*

We also tested QR reconstruction under the bright on bright settings, from a 40" dashboard display screen (Samsung UA40B6000VR) instead of our basic computer display. In this test, we used our entry-level DSLR camera (using a 35mm lens). For QR objects, reconstruction was 70-100% at distances of 150-350cm.

## 6. Scientific discussion

In this section we present the scientific background regarding human visual perception limitations which facilitate the success our optical covert channel. The ability of humans to resolve blinking images and brightness perception are discussed in [38] [39] [40] [41] [49]. Physiological aspects of human color vision are discussed further by Gouras [50]. Coren et al also provide a general discussion of the human visual system [51], along with details regarding the perception of brightness and darkness [52], lightness constancy [53], and temporal properties of the visual system [54].

### 6.1 Brightness and darkness perception

The level of ambient (environmental) light is known to affect visual perception, including the perception of brightness [52]. In fact, the human visual system consists of photopic or daylight vision, which includes the perception of color, and scotopic or twilight vision. In the human retina, two separate types of cells (cones and rods) are responsible for daylight and twilight vision: cones are associated with photopic vision, while rods are associated with scotopic vision. The sensitivity of the visual system gradually adapts as one move to a darker or brighter environment. Consequently, our experiments are performed under a controlled level of ambient light. Also, subjects are given some time to adapt to the laboratory's level of ambient light. It is also believed that human perception of relative brightness and darkness involves two separate systems [52].

### 6.2 Blinking image perception

Concerning the duration of the blinking image, particularly with low levels of illumination, increasing the duration can increase the likelihood that the stimulus will be detected, a phenomenon known as Bloch's law [52]. Concerning perception of flickering light, the retinal receptors in the human eye can resolve up to several hundred cycles per second (cps). However, the sensitivity of neurons in the primary visual cortex to flickers is much lower [54]. The critical fusion frequency (CFF) is used to measure subjects' discrimination between steady and flickering light. This measure varies between 10 cps and 60 cps (exposure time between 50 ms and 8.3 ms, respectively). The CFF varies based on several factors, including the current level of light/dark adaptation, the intensity of the light, the distance from the fovea, and the wavelength composition of the light. Consequently, our experiments are performed under controlled values for those factors. In the human retina, separate ganglion cells are responsible for sustained (steady) light and transient (flickering) light (see also [51]). Interestingly, it has been demonstrated [54] that low-contrast flashes and equiluminant chromatic (color) flashes activate different pathways. In this research, we are particularly interested in low-contrast flashes of gray tones.

## 7. Countermeasures

As with other kinds of emanation-based data leakage [9], countermeasures can be grossly categorized into procedural versus technological countermeasures.

**Procedural countermeasures** may include organizational practices aimed to restrict the accessibility of sensitive computers by placing them in secured spatial zones where only highly authorized staff may access them, or by barring any sort of cameras (including mobile phones and the like) from being carried within the perimeter of the secured zone. Such countermeasures, however, are not applicable in the case of dashboards, ATMs and similar machines, which are usually placed in public locations. In that case, the presence of surveillance cameras may have some deterring effect on potential shoulder surfers. However, the surveillance camera itself may be compromised by a malicious program [27].

**Technological countermeasures** may include scanning of the sensitive computer for the presence of suspicious display-oriented patterns or anomalies at runtime. Practical implementation of such countermeasures, however, appears to be nontrivial, since the covert channel by itself does not perform any explicitly network actions that would normally be detected as data leakage. Another possibility is taking periodical photos or videos of the computer's display and trying to scan them, searching for obscured visual patterns. Again, practical implementation seems nontrivial, unless one knows what to look for; in this case, anomaly detection techniques may be helpful. Another technological countermeasure consists of a thin polarized film which covers the display. The user, placed in front of the display, gets a clear view, while people around him/her would see a darkened display. This technology is commercially available [55] [56], for protecting the privacy of computer users. Its primary use is protecting portable computers or mobile phones from shoulder surfing in public places. However, this solution may be more challenging in the case of ATMs and similar machines, where the potential attacker can place himself in front of the display.

## 8. Conclusion and future work

This work introduces a new type of covert optical channel which, unlike existing optical methods (e.g., blinking LEDs [13]), can also be used while the user is working on the computer. We exploit the boundaries of human vision by projecting very close levels of bright and dark images on the computer's display. We also take advantage of the limits of the perception of flickering objects. As demonstrated in our research, using a digital camera combined with digital image processing, we were able to reconstruct objects which were not perceived by human subjects. The covert channel can be produced quite easily, using standard LCD displays and digital cameras. We conducted extensive tests with 40 users and various types of off-the-shelf cameras to examine the limitations of the covert channel. Notably, this kind of covert channel is not monitored by existing data leakage prevention systems. Future work should include further investigation of human vision limitations, particularly concerning complex colored backgrounds. Another direction is to use specialized optical lenses to extend the effective distance and reconstruction quality. Considering the growing popularity of wearable cameras and the continuous improvements made to the now ubiquitous cameras on mobile phones, the scope, effectiveness, and potential for harm of covert optical channels such as VisiSploit are expected to increase dramatically in the future. We hope that our work will raise interest in optical covert channels and promote further academic research in this domain.

**APPENDIX A**

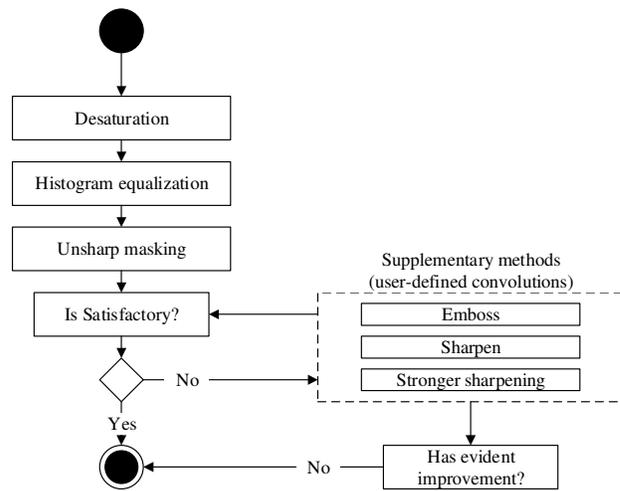

**Figure 8. Outline of the object reconstruction process.**

In Figure 8 we present the outline of the object reconstruction process, focusing on the main methods used at each step. The process starts with desaturation, proceeds to histogram equalization, and then continues on to unsharp masking. If the result is satisfactory at this phase, the process terminates. If the result is not satisfactory, the process turns to supplementary methods (user-defined convolutions), including emboss, sharpen, and stronger sharpening, as needed. The process finally terminates when the result is satisfactory or when there is no evident improvement. Note that this figure is oversimplified. The various steps and the aforementioned methods are further explained in the following subsections.

**Desaturation**

The first major step of almost any image processing task is removing various types of noise that are entangled with the original signal and would hamper its proper reconstruction. Different kinds of filters apply to different kinds of noise. Subsequently, one should determine the filters that are most effective with the given image. Noise-removal filters can be applied either in the spatial domain, using convolution methods, or in the frequency domain. In our case, the original image as produced by the digital camera is a color image, but the informative spectrum involves only grayscale. Consequently, we apply desaturation [57] [58], which removes the so-called chroma-noise.

**Histogram equalization**

In our experiments, the second major step involves sharpening the image, i.e., enhancing the contrast between the object and its background. Recall that we intentionally decreased the contrast to a level at which the object is unperceivable by a human subject. As with filtering, this step may involve the application of some convolution methods in the spatial domain. Histogram equalization [59] was used as the first contrast-enhancing tool.

## Unsharp masking

The second contrast-enhancing method we used was unsharp masking [60] [46]. Contrary to its name, unsharp masking is an image processing method used to sharpen an image. Convolution matrix [61] for unsharp masking (with no image mask) takes the following form:

$$\frac{-1}{256}\begin{bmatrix} 1 & 4 & 6 & 4 & 1 \\ 4 & 16 & 24 & 16 & 4 \\ 6 & 24 & -476 & 24 & 6 \\ 4 & 16 & 24 & 16 & 4 \\ 1 & 4 & 6 & 4 & 1 \end{bmatrix}$$

We applied GIMP's built-in unsharp masking method. The adjustable parameters for that method are the radius, amount, and threshold. We used rather high radius values (30-50) and amount values (3.5-5.0), and rather low threshold values (0-10, with 0 being the typical value).

## Supplementary processing

When the contrast level achieved by the previous methods was not satisfactory, we applied supplementary methods, using various convolution (or kernel) matrices [61]. GIMP allows user-defined convolution matrices. The optional convolution matrices used at this step are presented below.

| Emboss | Sharpen | Stronger Sharpening |
|---|---|---|
| $\begin{matrix} 3 & -1 & 0 \\ -1 & 1 & 1 \\ 0 & 1 & -3 \end{matrix}$ | $\begin{matrix} 0 & -1 & 0 \\ -1 & 5 & -1 \\ 0 & -1 & 0 \end{matrix}$ | $\begin{matrix} 0 & 0 & -1 & 0 & 0 \\ 0 & 0 & -2 & 0 & 0 \\ -1 & -2 & 13 & -2 & -1 \\ 0 & 0 & -2 & 0 & 0 \\ 0 & 0 & -1 & 0 & 0 \end{matrix}$ |

The emboss and sharpen convolution matrices, as shown above, are conventional, while the stronger sharpening convolution matrix was specially crafted and was used when the conventional matrix failed.